\documentclass[a4paper]{jpconf}
\usepackage{graphicx,epsfig}
\usepackage{amsmath,amssymb}
\usepackage{fix-cm}
\begin{document}

\title{Electric dipole moments from spontaneous CP violation in SU(3)-flavoured SUSY}

\author{J Jones~P\'erez}
\address{$^2$ Departament de F\'{\i}sica Te\`orica and IFIC, Universitat de Val\`encia-CSIC, E-46100, Burjassot, Spain.}
\ead{joel.jones@uv.es}

\begin{abstract}
The SUSY flavour problem is deeply related to the origin of flavour and hence to the origin of the SM Yukawa couplings themselves. Since all CP-violation in the SM is restricted to the flavour sector, it is possible that the SUSY CP problem is related to the origin of flavour as well. In this work, we present three variations of an $SU(3)$ flavour model with spontaneous CP violation. Such models explain the hierarchy in the fermion masses and mixings, and predict the structure of the flavoured soft SUSY breaking terms. In such a situation, both SUSY flavour and CP problems do not exist. We use electric dipole moments and lepton flavour violation processes to distinguish between these models, and place constraints on the SUSY parameter space.
\end{abstract}

\section{Introduction}

Although Supersymmetry (SUSY) has not been observed yet in nature, it has become one of the most popular theories beyond the Standard Model (SM). This popularity is justified by the way it provides solutions to theoretical and cosmological problems that the SM cannot explain. Nonetheless, the Minimal Supersymmetric Standard Model (MSSM), has many problems of its own. Among these problems, two of the most important are:

\begin{description}
 \item [The SUSY Flavour Problem:] As we do not have any information about the off-diagonal elements of the soft SUSY-breaking matrices, we could expect them to be all of the same order. However, this generally gives too-large contributions to flavour changing neutral currents (FCNC) and unobserved processes like Lepton Flavour Violation (LFV). Thus, the off-diagonal elements must be suppressed with respect to the diagonal ones.

 \item [The SUSY CP Problem:] Among the many parameters of the MSSM, we have new CP violating (CPV) phases. It turns out that the phases associated with flavour independent parameters give a too-large contribution to the electron and neutron electric dipole moments (EDMs). This requires these phases to be suppressed if the masses of the SUSY partners are light.
\end{description}

Presented in this way, the suppressions suggest the existance of a naturalness problem within low-energy SUSY. Nonetheless, the origin of these problems, and their possible solutions, may be understood by observing the SM itself. In particular, the SM also presents its own Flavour Problem: if all elements in the flavoured matrices of the Standard Model (the Yukawa matrices) are assumed to be of the same order, one would not be able to reproduce the observed hierarchy in the mass spectrum and mixing angles. Thus, the flavoured matrices in the SM need to have a structure. It can then be said that the real Flavour Problem lies in the inability to understand the generation of structures in flavoured parameters, may these be in the form of Yukawa couplings or soft SUSY-breaking terms.

It is interesting to see that the flavoured elements of both the SM and the MSSM need complicated structures. This suggests that all flavoured parameters could have a common origin, and that by solving the Flavour Problem of the SM it might be possible to find a way to solve the Flavour Problem in the MSSM.

On the other hand, there is no such analogy for the SUSY CP Problem in the SM. Nevertheless, it is crucial to take into account that, although the CKM phase is large, all CPV in the SM comes from supressed flavour-dependent terms. In the MSSM, the problematic phases come from large flavour-independent terms. This might suggest that CPV is not completely understood, and that it should be constrained within the flavour sector.

An interesting solution for both of these problems lies in the use of family symmetries~\cite{Froggatt:1978nt}. The breaking of such a family symmetry can be used to give shape to both Yukawa matrices and soft terms. Moreover, this breaking of the family symmetry can also be associated with the spontaneous breaking of an exact CP symmetry~\cite{Ross:2004qn}, constraining in this way all CPV within the flavour sector. Furthermore, since the model predicts a structure for the soft terms, it is possible to estimate the order of magnitude of the contributions to flavoured observables, being able to reject the model or not from the low energy phenomenology.

In this work, a model based on an $SU(3)$ family symmetry will be presented~\cite{Ross:2004qn}. After describing the possible variations to this model, low energy phenomenology in the leptonic sector will be analyzed. In particular, the expectations for LFV and EDM observables shall be given.

\section{The SU(3) Model}

An $SU(3)$ flavour symmetry can explain the peculiar structure of the SM Yukawa couplings as the result of spontaneous symmetry breaking~\cite{King:2001uz}. The three generations of each SM fermionic field are grouped in an $SU(3)$ triplet \textbf{3}, such that the SM Yukawa couplings are not allowed in the limit of exact symmetry. In order to generate the Yukawas, the symmetry is broken by one or several scalar vacuum expectation values (vevs) of new fields, called flavons. The flavons are coupled to the SM fields through non-renormalizable operators, suppressed by a heavy mediator mass, to compensate the $SU(3)$ charges. If the scalar vev is smaller than the mediator scale, this provides a small expansion parameter that can be used to explain the hierarchy of the observed Yukawa couplings.

In the same way, these flavons will couple to the SUSY scalar fields in all possible ways allowed by the symmetry and, after spontaneous symmetry breaking, they will generate a non-trivial flavour structure in the soft-breaking parameters. Therefore, by being generated by insertions of the same flavon vevs, the structures in the soft-breaking matrices and the Yukawa couplings are related. The starting point in the analysis of the soft-breaking terms must then necessarily involve an analysis of the texture in the Yukawas, in order to reproduce first the correct masses and mixings.

Two assumptions are required to fix the Yukawa couplings: (1) the smallness of CKM mixing angles is due to the smallness of the off-diagonal elements in the Yukawa matrices with respect to the corresponding diagonal elements, and (2) the matrices are symmetric (for simplicity). With these two theoretical assumptions, and using the ratio of masses at a high scale to define the expansion parameters in the up and down sector as $\bar\varepsilon=\sqrt{m_s/m_b}\approx0.15$ and $\varepsilon= \sqrt{m_c/m_t}\approx0.05$, the Yukawa textures in the quark sector can be fixed to~\cite{Roberts:2001zy}:
\begin{eqnarray}  
\label{fit}
Y_d\propto\left( 
\begin{array}{ccc}
0 & x_{12} \ \bar \varepsilon^{3} &x_{13}  {\ \bar \varepsilon^{3}} \\ 
x_{12}\ \bar \varepsilon^{3} & {\bar \varepsilon^{2}} & 
x_{23}\ \bar \varepsilon^{2} \\ 
x_{13}\ \bar \varepsilon^{3} & x_{23} \ \bar \varepsilon^{2} & 1%
\end{array}%
\right),~~~~~~ Y_u\propto \left( 
\begin{array}{ccc}
0 & b^\prime {\ \varepsilon^{3}} & c^\prime {\ \varepsilon^{3}} \\ 
 b^\prime{\ \varepsilon^{3}} & {\ \varepsilon^{2}} & 
a^\prime \varepsilon^{2} \\ 
 c^\prime{\ \varepsilon^{3}} & a^\prime \varepsilon^{2} & 1%
\end{array}
\right ) \, ,
\end{eqnarray}
with $x_{12}=1.6$, $x_{13}=0.4$, $x_{23}=1.75$,  and $a^\prime,b^\prime,c^\prime$ being poorly fixed from experimental data. On the other hand, the Yukawa couplings in the leptonic sector can not be determined from the available phenomenological data. Within a seesaw scenario, the left-handed neutrino masses and mixings do not give any information to fix the neutrino Yukawa couplings. Therefore only the charged lepton masses provide useful information on leptonic Yukawas. For simplicity, Yukawa unification at high scales is assumed, a possibility favouring Grand Unified models. In this case, charged lepton and down-quark (and the neutrino and up-quark) Yukawa matrices are the same except for the different vev of an additional Georgi-Jarlskog Higgs field $\Sigma$, in the $(B-L+2T_3^R)$ direction, to unify the second and first generation masses.

The $SU(3)$ model has several flavon fields, denoted $\theta_3$, $\theta_{23}$ (anti-triplets ${\bf \bar 3}$), $\bar \theta_3$ and $\bar \theta_{23}$ (triplets ${\bf 3}$). These have to be coupled to the matter fields using effective couplings in every possible way. Unfortunately, given a particular symmetry breaking pattern for the flavons, the $SU(3)$ symmetry is not restrictive enough to reproduce the textures in Eq.~(\ref{fit}) and one must impose some additional symmetries to guarantee the correct power structure and to forbid unwanted terms in the effective superpotential. The basic structure of the Yukawa superpotential (for quarks and leptons) is then given by:
\begin{eqnarray}
W_{\rm Y} &=& H\psi _{i}\psi _{j}^{c} \left[\theta _{3}^{i}  \theta
_{3}^{j}+\theta _{23}^{i} \theta _{23}^{j}
\left(\theta_3\overline{\theta}_3\right)\Sigma  + 
\epsilon ^{ikl} \overline{ \theta}_{23,k} {\overline{ \theta }_{3,l}} \theta _{23}^{j}\left(
 \theta _{23} {\overline{\theta} _{3}}\right) +\dots \right],
\end{eqnarray} 
where to simplify the notation, the flavon and $\Sigma$ fields have been normalized to the corresponding mediator mass, i.e., all the flavon fields in this equation should be understood  as $\theta_i/M_f$.  

After spontaneous breaking of the flavour symmetry (and CP symmetry) the vevs
of the different flavon fields are:
\begin{align}
\label{vevs}
\langle\theta_3\rangle=\left( 
\begin{array}{c}
0 \\ 
0 \\ 
1\end{array}
\right)&\otimes \left(
\begin{array}{cc}
a_3^u & 0 \\ 
0 & a_3^d~ e^{i \chi}
\end{array}
\right);& \langle\bar{\theta}_3\rangle=\left( 
\begin{array}{c}
0 \\ 
0 \\ 
1\end{array}
\right)&\otimes \left(
\begin{array}{cc}
a_3^u~ e^{i \alpha_u} & 0 \\ 
0 & a_3^d~ e^{i \alpha_d}
\end{array}
\right);  \nonumber \\
\langle\theta_{23}\rangle=&\left( 
\begin{array}{c}
0 \\ 
b_{23} \\ 
b_{23}~e^{i\beta_3}
\end{array}
\right);& \langle\bar\theta_{23}\rangle=&\left( 
\begin{array}{c}
0 \\ 
b_{23}~e^{i\beta^{\prime}_2} \\ 
b_{23}~e^{i(\beta^{\prime}_2 - \beta_3)}%
\end{array}
\right);
\end{align}
where a vacuum alignment mechanism is required. The following relations are needed to properly reproduce the Yukawas in Eq.~(\ref{fit}):
\begin{align}
\label{expansion}
\left(\frac{\displaystyle{a_3^u}}{\displaystyle{M_u}}\right)^2& = y_t,&
\left(\frac{\displaystyle{a_3^d}}{\displaystyle{M_d}}\right)^2& = y_b, \nonumber \\
\frac{\displaystyle{b_{23}}}{\displaystyle{M_u}}& = \varepsilon,& \frac{\displaystyle{b_{23}}}{\displaystyle{M_d}}& = \bar \varepsilon.
\end{align}

This structure is quite general for the different $SU(3)$ models one can build, and for additional details we refer to \cite{Ross:2004qn,King:2001uz,King:2003rf}. It is important to notice that the flavon vevs in Eq.~(\ref{vevs}) carry phases. This means that if exact CP symmetry is demanded in the original lagrangian, the breaking of $SU(3)$ also breaks CP invariance. By doing this, one can restrict all CPV to lie within the flavour sector, removing the large flavour-independent phases that are the source of the SUSY CP Problem. It has also been shown in~\cite{Ross:2004qn} that it is possible to reproduce the observed CKM phase using the flavon phases. This procedure shall be followed in this work.

As mentioned previously, in the context of a supersymmetric theory an unbroken flavour symmetry would apply equally to the fermion and scalar sectors. This implies that in the limit of exact symmetry the soft-breaking scalar masses and the trilinear couplings must also be invariant under the flavour symmetry. This has different implications in the case of the scalar masses and the trilinear couplings, although in the present work only the scalar masses will be analyzed (we will take $A_0=0$).

Scalar flavour-diagonal couplings are invariant under any symmetry, which means that diagonal soft-masses are always allowed by the flavour symmetry and will be of the order of the SUSY breaking scale. Additionally, for $SU(3)$, the three generations are grouped in a single multiplet with a common mass, thus reducing greatly the FCNC problem.

After $SU(3)$ breaking, the scalar soft masses deviate from exact universality, and any invariant combination of flavon fields contribute to the sfermion masses. In this case, the following terms will always contribute to the sfermion mass matrices:
\begin{align}
\label{eq:minimal}
(M^2_{\tilde f})^{ij} = m_0^2 \bigg(\delta ^{ij} &
  +\frac{\displaystyle{1}}{\displaystyle{M_f^{2}}}\left[\theta _{3, i}^{\dagger}
\theta _{3,j}  + \overline{ \theta} _{3}^{i\dagger } \overline{ \theta} _{3}^{j} + 
\theta _{23, i}^{\dagger }\theta_{23,j} + \overline{ \theta} _{23}^{i\dagger } 
\overline{ \theta}_{23}^{j} \right] \nonumber \\
 & + \frac{1}{M_f^4}(\epsilon ^{ikl}\overline{\theta }_{3,k}
\overline{\theta }_{23,l})^{\dagger }(\epsilon ^{jmn}
\overline{\theta }_{3,m}\overline{\theta }_{23,n})+\quad\ldots\quad\bigg),
\end{align}
where $f$ represents the $SU(2)$ quark and lepton doublets or the up (neutrino) and down (charged-lepton) singlets. Notice there are three different mediator masses, $M_f=M_L, M_u, M_d$, although for simplicity $M_u$ is taken equal to $M_L$. This minimal structure shall be denoted as RVV1. The soft scalar masses of the slepton sector, in the SCKM basis, take the following structure:
\begin{subequations}
\label{sckm1}
\begin{eqnarray}
\frac{(m^2_{\tilde{e}^c_R})^T}{m_0^2} & = & \left(\begin{array}{ccc}
1+\bar\varepsilon^{2}~y_d & \frac{1}{3}\bar\varepsilon^3 & \frac{1}{3}\bar\varepsilon^3~e^{-i(\beta_3-\chi)} \\
\frac{1}{3}\bar\varepsilon^3 & 1+\bar\varepsilon^2 & \bar\varepsilon^2~e^{-i(\beta_3-\chi)} \\
\frac{1}{3}\bar\varepsilon^3~e^{i(\beta_3-\chi)} & \bar\varepsilon^2~e^{i(\beta_3-\chi)} & 1+y_d\end{array}\right) \\
\frac{m^2_{\tilde{L}}}{m_0^2} & = & \left(\begin{array}{ccc}
1+\varepsilon^2~y_t & \frac{1}{3}\varepsilon^2\bar\varepsilon & \bar\varepsilon^{3}~y_t ~e^{-i(\beta_3-\chi)} \\
\frac{1}{3}\varepsilon^2\bar\varepsilon & 1+\varepsilon^2 &
3 \bar\varepsilon^{2}~y_t ~e^{-i(\beta_3-\chi)} \\
\bar\varepsilon^{3}~y_t ~e^{i(\beta_3-\chi)} &
3 \bar\varepsilon^{2}~y_t ~e^{i(\beta_3-\chi)} & 1+ y_t
\end{array}\right)
\end{eqnarray}
\end{subequations}
where for simplicity we have neglected $O(1)$ constants, which usually have important subleading phases.

It is possible to build other invariant combinations with different flavon fields that can not be present in the superpotential. This is due to the fact that the superpotential must be holomorphic, i.e. can not include daggered fields, while the soft masses, coming from the K\"ahler potential, only need to be real combinations of fields. The type of allowed extra terms depend on the symmetries one imposes to shape the Yukawa matrices. Such symmetries can allow two exclusive variations without modifying the initial Yukawa structures.

The first variation to the minimal structure is achieved by allowing a $\theta_3^i\bar\theta_{23}^j$ term in the soft mass matrix (RVV2). When rotated to the SCKM basis, the mass matrices become:
\begin{subequations}
\label{sckm2}
\begin{eqnarray}
\frac{(m_{\tilde e_R^c}^2)^T}{m^2_0} & = & \left(\begin{array}{ccc}
1+\bar\varepsilon^2\,y_b & \frac{1}{3}\bar\varepsilon^3 & \frac{1}{3}\bar\varepsilon^2\,y_b^{0.5}\,e^{-i\beta'_2} \\
\frac{1}{3}\bar\varepsilon^3 & 1+\bar\varepsilon^2 & \bar\varepsilon\,y_b^{0.5}\,e^{-i\beta'_2} \\
\frac{1}{3}\bar\varepsilon^2\,y_b^{0.5}\,e^{i\beta'_2} & \bar\varepsilon\,y_b^{0.5}\,e^{i\beta'_2} & 1+y_b
\end{array}\right) \\
\frac{m_{\tilde L}^2}{m^2_0} & = & \left(\begin{array}{ccc}
1+\varepsilon^2\,y_t & \frac{1}{3}\varepsilon^2\bar\varepsilon & \frac{1}{3}\varepsilon\bar\varepsilon\,y_t^{0.5}\,e^{-i(\chi-\beta'_2)} \\
\frac{1}{3}\varepsilon^2\bar\varepsilon & 1+\varepsilon^2 & \varepsilon\,y_t^{0.5}\,e^{-i(\chi-\beta'_2)} \\
\frac{1}{3}\varepsilon\bar\varepsilon\,y_t^{0.5}\,e^{i(\chi-\beta'_2)} & \varepsilon\,y_t^{0.5}\,e^{i(\chi-\beta'_2)} & 1+y_t
\end{array}\right)
\end{eqnarray}
\end{subequations} 
Using mass-insertion notation~\cite{Gabbiani:1996hi, Hagelin:1992tc}, one can see that the effect of this term in $(m_{\tilde e_R^c}^2)^T$ is to exchange one power of $\bar\varepsilon$ by a $y_b^{0.5}$ supression in $(\delta^e_{13})_{RR}$ and $(\delta^e_{23})_{RR}$. In $m_{\tilde L}^2$, the same terms change an $\bar\varepsilon^2$ by an $\varepsilon\,y_t^{0.5}$. However, for $\tan\beta=10$, and considering that $\varepsilon\approx\bar\varepsilon^2$, such replacements leave the structure of the mass matrices very similar numerically to the original one. Nonetheless, it must be remarked that the phase structure of the whole mass matrix is modified.

The second variation to the minimal structure allows a $\left(\epsilon^{ikl}\theta_3^k\theta_{23}^l\right)\theta_3^j$ term (RVV3). The soft matrices, when rotated into the SCKM basis, have the following structure:
\begin{subequations}
\label{sckm3}
\begin{eqnarray}
\frac{(M_{\tilde e_R^c}^2)^T}{m_0^2} & = & \left(\begin{array}{ccc}
1+\bar\varepsilon^2\,y_b & \frac{1}{3}\bar\varepsilon^3 & \bar\varepsilon\,y_b\,e^{-i(\delta_d-\chi)} \\
\frac{1}{3}\bar\varepsilon^3 & 1+\bar\varepsilon^2 & \bar\varepsilon^2\,e^{-i(\beta_3-\chi)} \\
\bar\varepsilon\,y_b\,e^{i(\delta_d-\chi)} & \bar\varepsilon^2\,e^{i(\beta_3-\chi)} & 1+y_b
\end{array}\right)\\
\frac{M_{\tilde L}^2}{m_0^2} & = & \left(\begin{array}{ccc}
1+\varepsilon^2\,y_t & 3\varepsilon\bar\varepsilon^2\,y_t\,e^{-i(2\chi-\beta_3-\delta_d)} & \varepsilon\,y_t\,e^{-i(\chi-\delta_d)} \\
3\varepsilon\bar\varepsilon^2\,y_t\,e^{i(2\chi-\beta_3-\delta_d)} & 1+\varepsilon^2 & 3\bar\varepsilon^2\,y_t\,e^{-i(\beta_3-\chi)} \\
\varepsilon\,y_t\,e^{i(\chi-\delta_d)} & 3\bar\varepsilon^2\,y_t\,e^{i(\beta_3-\chi)} & 1+y_t
\end{array}\right)
\end{eqnarray}
\end{subequations}
where $\delta_d=2\alpha_d+\beta'_2+\beta_3$.

This model shows larger deviations from RVV1 in the $LL$ sector. It is important to notice that $(\delta^e_{12})_{LL}$ is now of order $\varepsilon\bar\varepsilon^2$ instead of $\varepsilon^2\bar\varepsilon$, which will have considerable consequences in processes such as $\mu\to e\gamma$. Likewise, $(\delta^e_{13})_{LL}$ is of order $\varepsilon\,y_t$ instead of $\bar\varepsilon^3$, so an enhancement in $\tau\to e\gamma$ should be expected. Regarding the $RR$ sector, for $\tan\beta=10$, the $y_b$ suppression at $M_{GUT}$ has roughly the same size as an $\bar\varepsilon$ suppression, so once again the structure of $(m_{\tilde e_R^c}^2)^T$ is numerically similar to RVV1.

It is important to emphasize at this point that these deviations from universality in the soft-mass matrices due to flavour symmetry breaking come always through corrections in the K\"ahler potential. Therefore, these effects will be important only in gravity-mediation SUSY models where the low-energy soft-mass matrices are mainly generated through the K\"ahler potential. In other mediation mechanism, as gauge-mediation or anomaly mediation, where these K\"ahler contributions to the soft masses are negligible, flavour effects in the soft mass matrices will be basically absent.

The soft mass matrices shown are given at $M_{GUT}$, which means that no effects coming from the running have been included. Such running effects can be sizeable in the LL and LR sectors, due to the presence of heavy RH neutrinos with large Yukawa couplings \cite{Borzumati:1986qx,Casas:2001sr,Masiero:2002jn}. Moreover, in the present case there are new contributions to the running given by the non-universality of the soft mass matrices. These effects can be important even in the RR sector. Nevertheless, it turns out that these contributions are, at most, of the same order as the initial matrices. This means that the addition of running effects will only change the already unknown $O(1)$ constants, such that the low-energy phenomenology can still be understood by analyzing Eqs.~(\ref{sckm1})-(\ref{sckm3}).

\section{Phenomenology in the Leptonic Sector}

\subsection{Lepton Flavour Violation}

Supersymmetric flavour models characterized by unaligned, non-universal scalar masses imply the arising of potentially large mixing among flavours. This has an immediate impact on processes with FCNC and LFV, via loop diagrams. This means that the proposed $SU(3)$ family symmetry can be tested through such processes.

In particular, in the lepton sector, the mass insertions are sources of lepton flavour violation, through neutralino or chargino loop diagrams. As a consequence, it is expected to have constraints in the allowed SUSY parameter space due to the experimental limits on LFV decays such as $\mu \to e \gamma$~\cite{Calibbi:2008qt}.

\begin{figure}
\includegraphics[scale=.25]{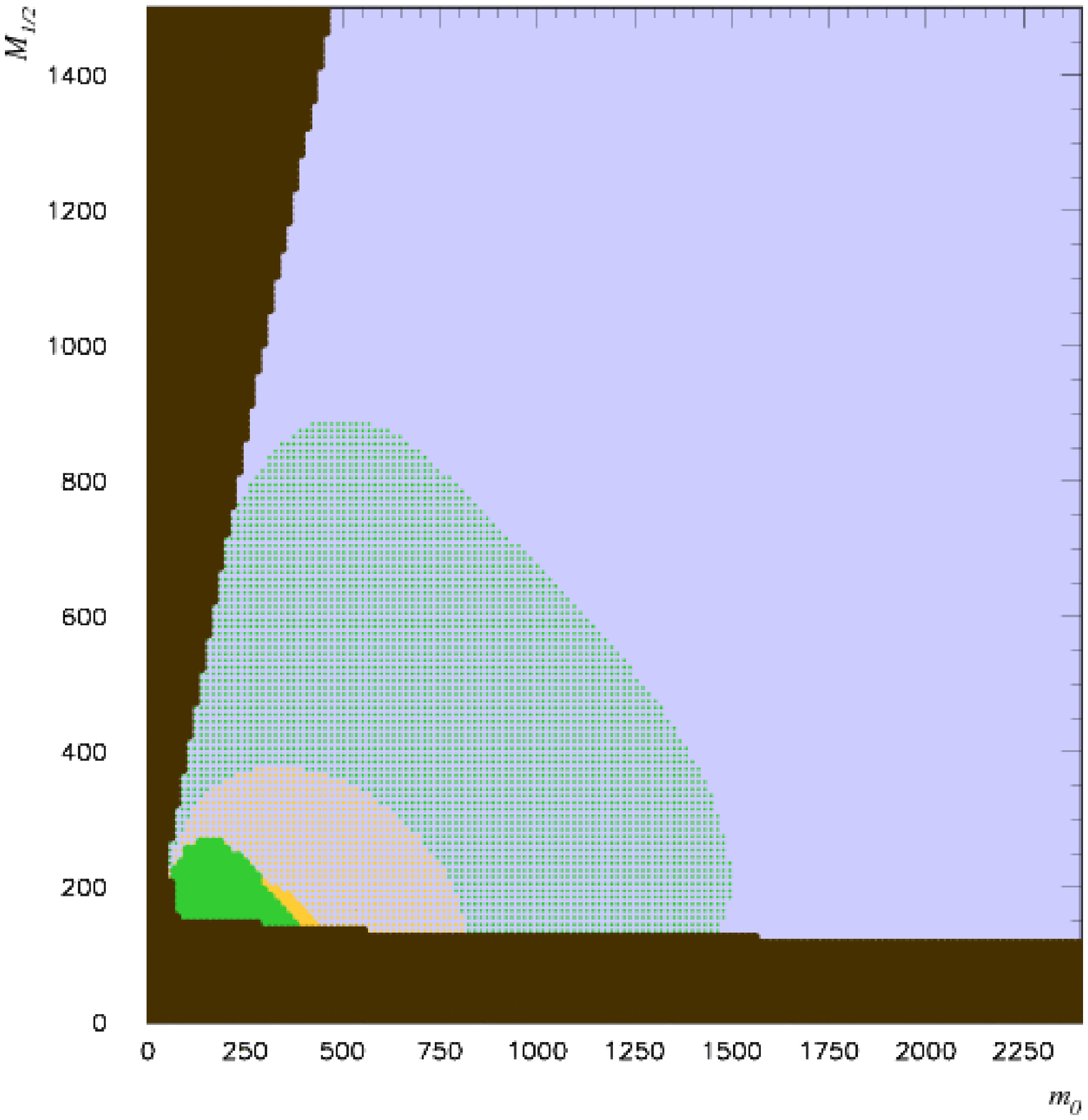} \includegraphics[scale=.25]{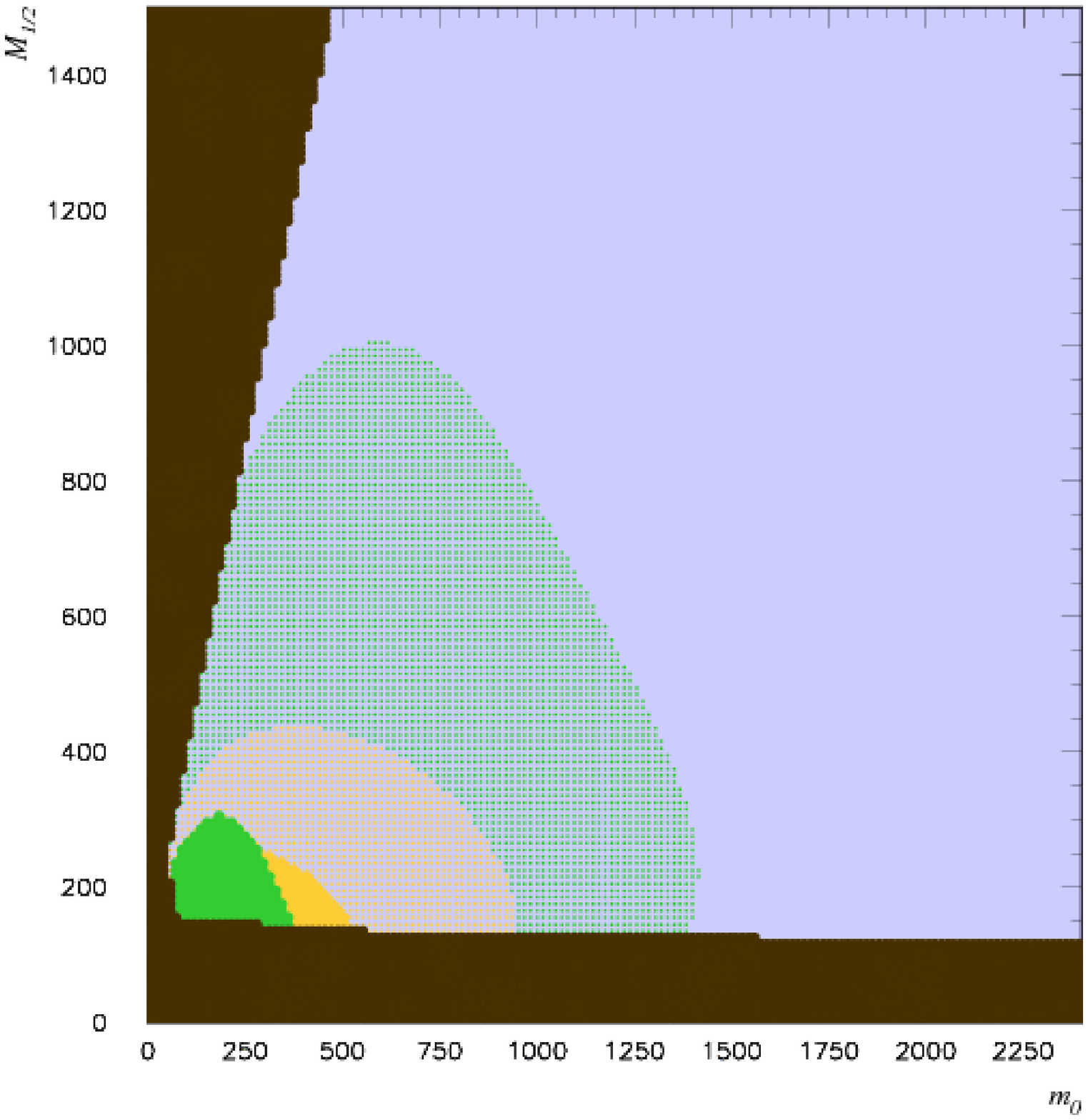} \includegraphics[scale=.25]{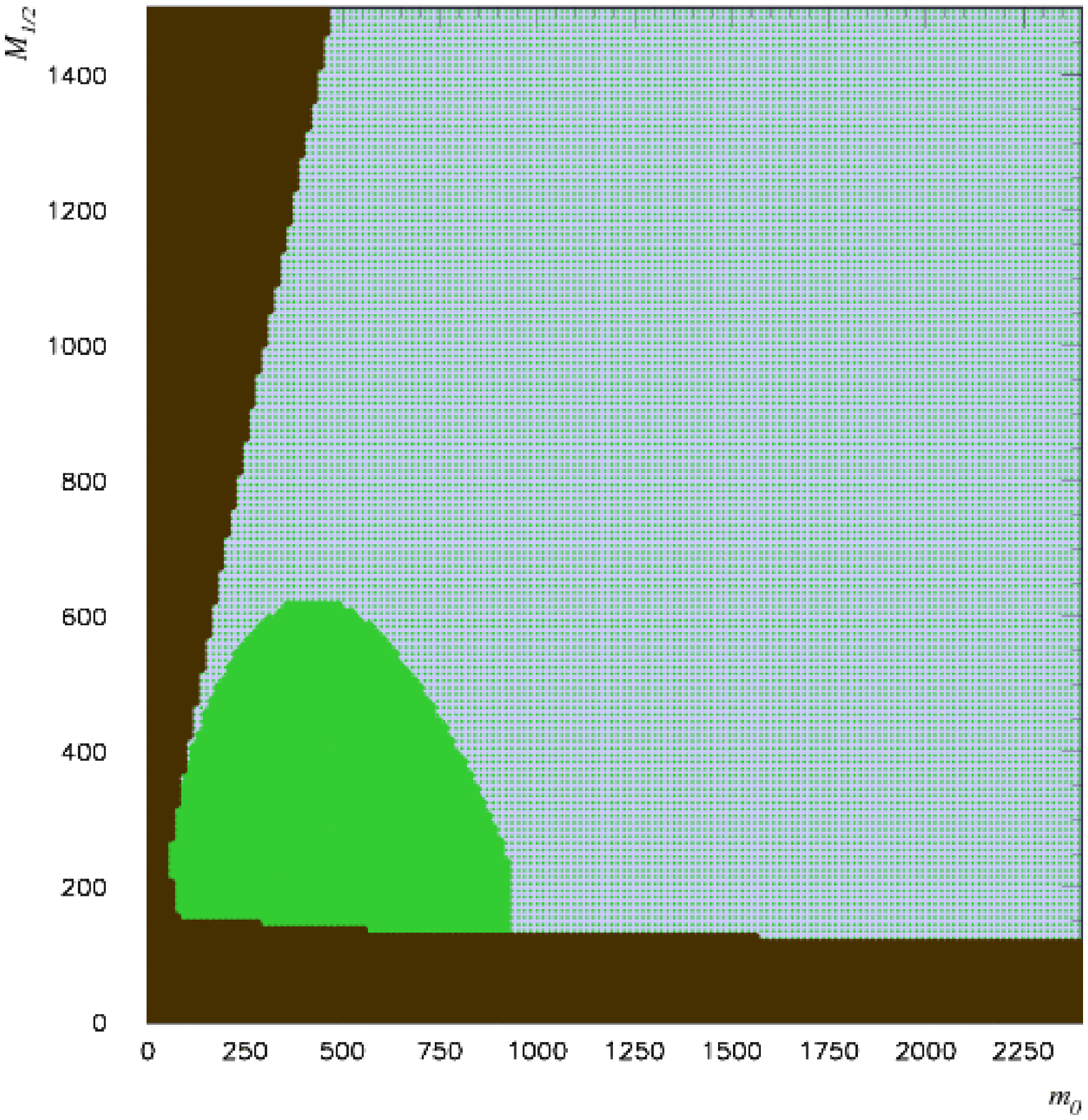}
\caption{Current constraints due to $\mu\to e\gamma$ (green) and $\tau\to\mu\gamma$ (yellow) in the $m_0$-$M_{1/2}$ plane for $\tan\beta =10$ and $A_0=0$. We show constraints for RVV1 (left), RVV2 (center) and RVV3 (right). The green dotted region corresponds to the reach of $\mu\to e \gamma$ at the MEG experiment, while the yellow hatched region is the reach of $\tau\to\mu\gamma$ at the Super Flavour Factory. Direct LEP bounds and charged LSP constraints are shown in dark brown.}
\label{fig:mueg}
\end{figure}

By setting $A_0=0$, the branching ratio of $\mu \to e \gamma$ decay depends only on the square of the $(\delta^e_{LL})_{21}$ and $(\delta^e_{RR})_{21}$ mass insertions. Inspecting the structures shown in Eqs.~(\ref{sckm1}), (\ref{sckm2}) and (\ref{sckm3}), it is evident that the $\mu\to e\gamma$ branching ratio will be similar for RVV1 and RVV2, while for RVV3 the $LL$ contribution will be enhanced by a factor $\bar\varepsilon/\varepsilon=3$. Since it is the square of the mass insertion what contributes to the branching ratio, we can expect an increase in the $LL$ contribution by an order of magnitude.

For $\tau\to\mu\gamma$, the difference between RVV1 and RVV2 is of a factor $\bar\varepsilon/y_b^{0.5}$ and $3(\bar\varepsilon^2/\varepsilon)y_t^{0.5}$ for the $RR$ and $LL$ contributions, respectively. Taking the values of the Yukawa couplings at $M_{GUT}$ for $\tan\beta=10$, one can check that both ratios are of order 1. Thus, no great differences should be expected either between RVV1 and RVV2. RVV3 has got the same sort of insertions as RVV1, which means that $\tau\to\mu\gamma$ decay is not sensitive to any of the possible variations.

The decay $\tau\to e\gamma$ has been analyzed in~\cite{Calibbi:2008qt}, and has been shown not to be sensitive enough to constrain these kind of models, in comparison with $\mu\to e \gamma$. Thus, by looking at the modulus of the mass-insertions, one can only distinguish RVV3 from RVV1 and RVV2.

An analysis of these three variations is shown in Figure~\ref{fig:mueg}, for $\tan\beta=10$ and $A_0=0$. The predictions for RVV1 and RVV2 are similar, as expected. Both models are slightly constrained by both $\mu\to e \gamma$~\cite{Ahmed:2001eh} and $\tau\to\mu\gamma$~\cite{Banerjee:2007rj}, where both branching ratios exclude SUSY masses with $m_0\lesssim250-400$~GeV, $M_{1/2}\lesssim200-300$~GeV. RVV3 is much more constrained, as $\mu\to e\gamma$ can be an order of magnitude larger. SUSY masses following $m_0\lesssim500-900$~GeV, $M_{1/2}\lesssim400-600$~GeV are excluded.

The future observation of $\mu\to e\gamma$ is crucial for all three models. The proposed increase in sensitivity by two orders of magnitude in the MEG experiment~\cite{MEG} will probe much of the parameter space accesible to the LHC. Thus, if SUSY is found at the LHC and any of these models is at work, it is very likely to observe $\mu\to e \gamma$ at MEG. The same can be said for $\tau\to\mu\gamma$ at the Super Flavour Factory~\cite{Bona:2007qt}, with an increase of sensitivity of one order of magnitude. This is also shown in Figure~\ref{fig:mueg}, although such constraints are not as strong.

\subsection{Electric Dipole Moments}

\begin{figure}
\includegraphics[scale=.25]{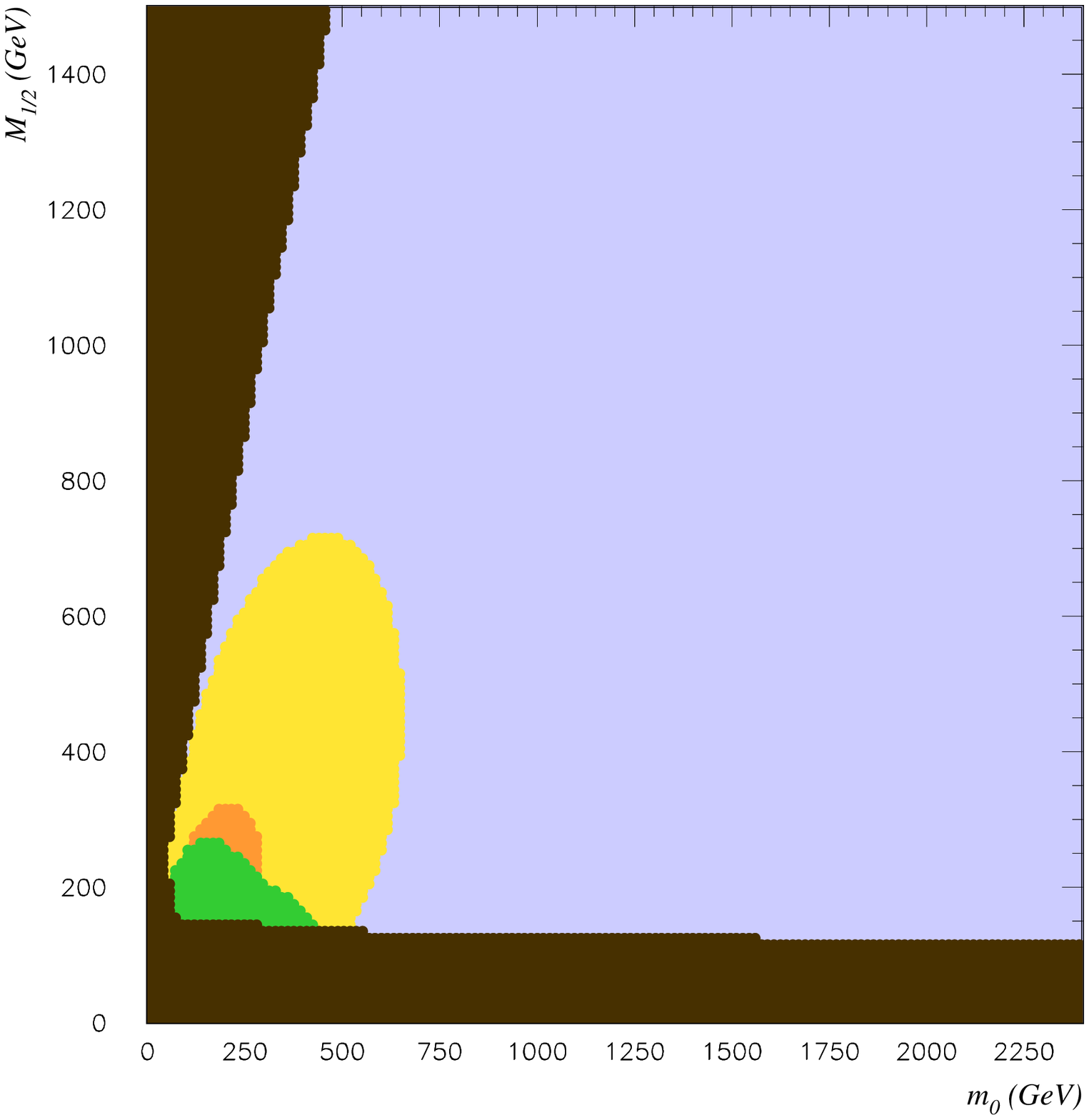} \includegraphics[scale=.25]{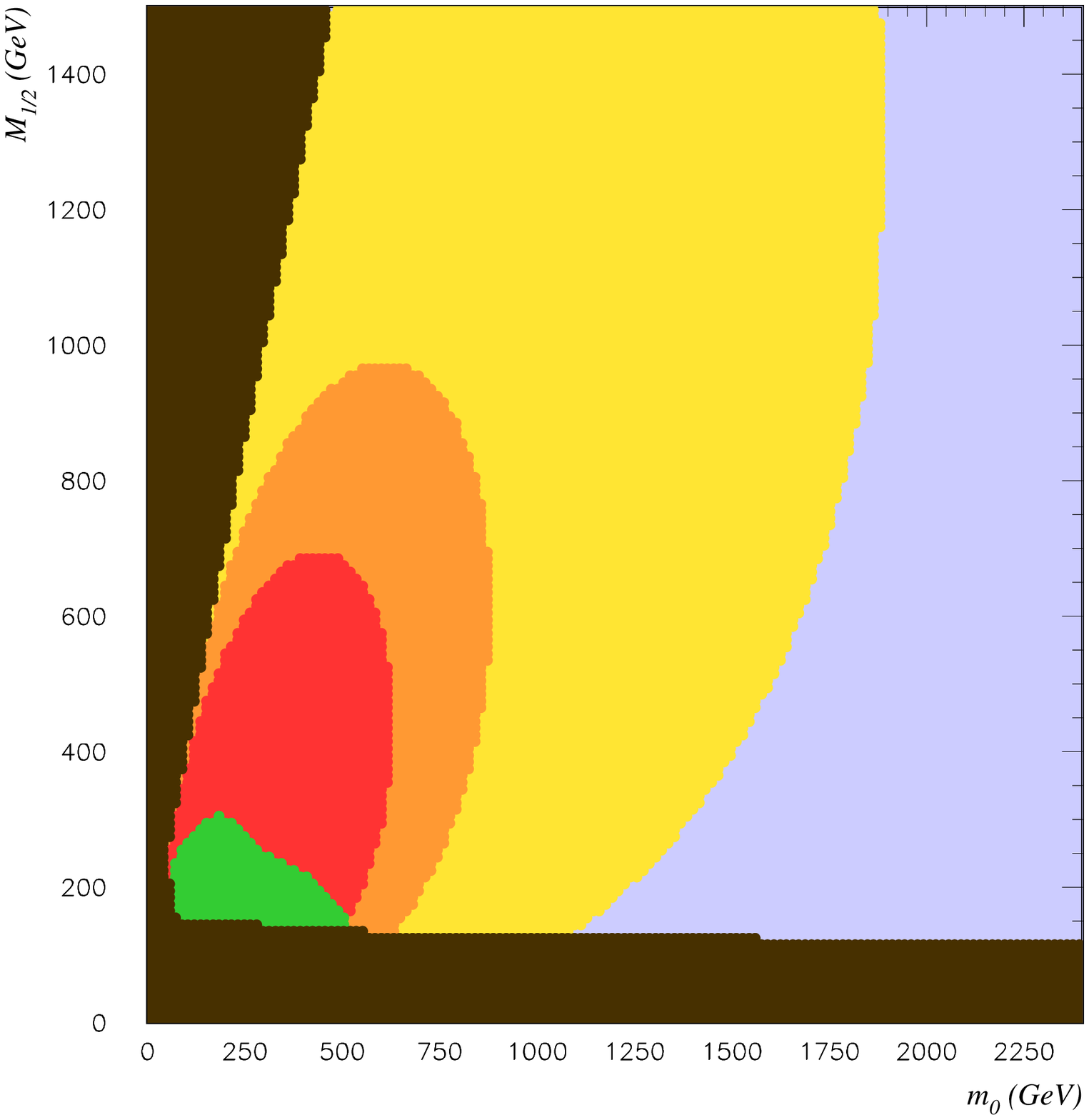} \includegraphics[scale=.25]{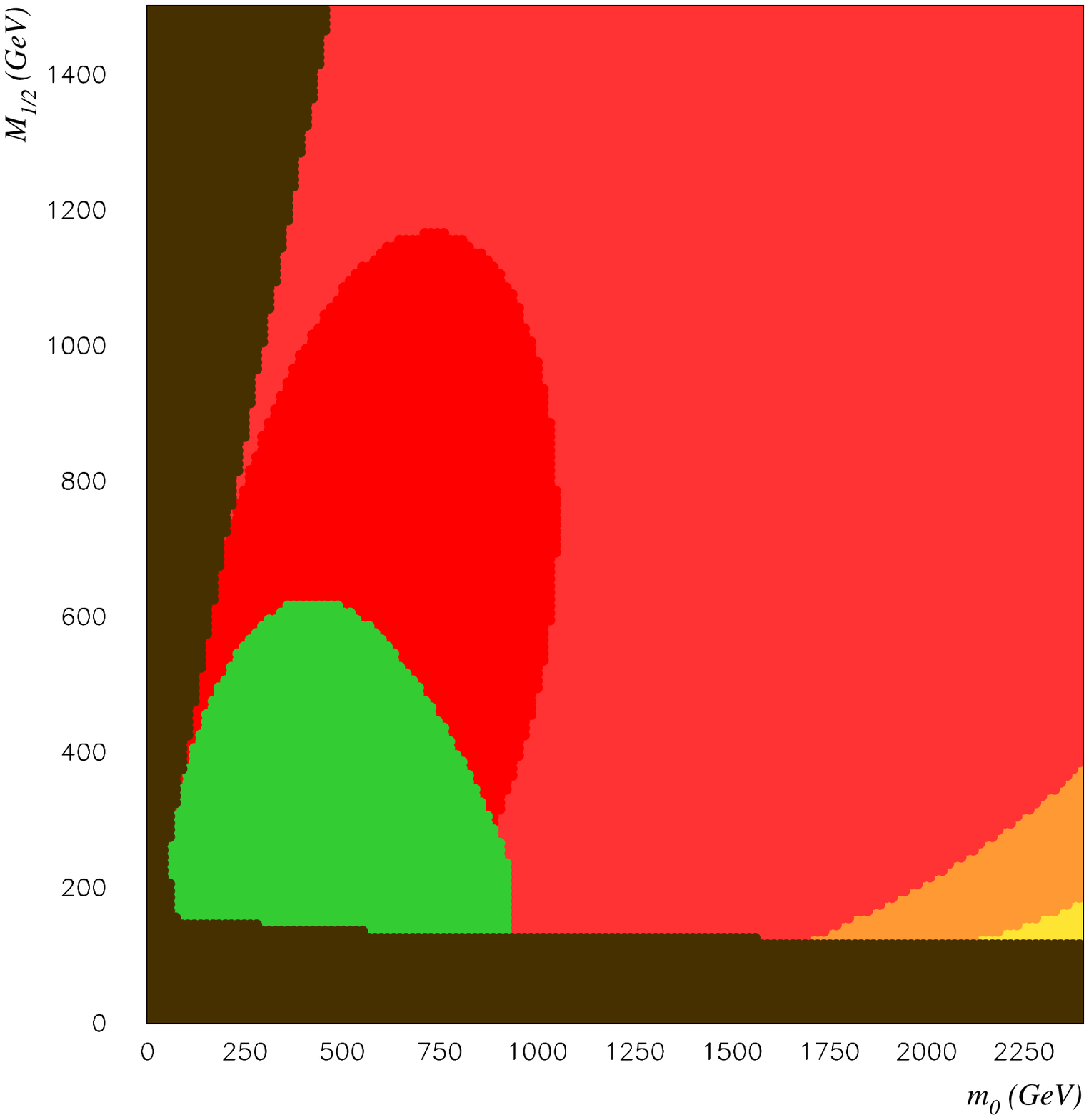}
\caption{Contours of $|d_e|=1\times10^{-28}$ e cm (light red), $|d_e|=5\times10^{-29}$ e cm (orange) and $|d_e|=1\times10^{-29}$ e cm (yellow) in the $m_0$-$M_{1/2}$ plane for $\tan\beta =10$ and $A_0=0$. We show predictions for RVV1 (left), RVV2 (center) and RVV3 (right). Current EDM bound ($1.4\times10^{-27}$) is shown in dark red. Current LFV bounds are also shown in green.}
\label{fig:edms}
\end{figure}

Even if $\mu\to e \gamma$ is observed at MEG, the question remains on how to distinguish between RVV1 and RVV2. The answer to this question lies on the observation of the electron EDM, $d_e$.

Since the $SU(3)$ models start with exact CP symmetry, broken by the complex flavon vevs, the problematic flavour-independent phases are gone. Nonetheless, $d_e$ can be still generated by combinations of phases in flavour violating terms. The three RVV models predict their own phase structure along with their flavour structure, so it is plausible to consider $d_e$ as a suitable observable for distinguishing between the models.

In~\cite{Calibbi:2008qt} it was shown that the most important contribution to $d_e$, when $A_0=0$, comes from a bino-mediated diagram proportional to $\Im m\left[(\delta^e_{13})_{LL}(\delta^e_{33})_{LR}(\delta^e_{31})_{RR}\right]$, which is enhanced by $m_\tau\tan\beta$. Due to the initial CP symmetry, the $(\delta^e_{33})_{LR}$ term is real, so one has to analyze the phases in the flavour-violating terms.

For RVV1, one can see in Eq.~(\ref{sckm1}) that the leading phases cancel. This means that $d_e$ will come from phases in subleading terms, contained within the $O(1)$ parameters, and thus will be highly supressed. The most important phase within these subleading terms depends on $2(\chi-\beta_3)$.

RVV2, even though having roughly the same order of magnitude for $\tan\beta=10$, has a non-vanishing leading phase. We can thus expect $d_e$ in RVV2 to be somewhat larger than in RVV1. Nonetheless, the largest $d_e$ comes from RVV3, which not only has a larger flavour structure, but also a leading non-vanishing phase.

Inspection of Eqs~(\ref{sckm1})-(\ref{sckm3}) leads to the following predictions:
\begin{eqnarray}
 (d_e)_{\textnormal{RVV1}} & \ll& \bar\varepsilon^6\frac{y_t}{3} \\
 (d_e)_{\textnormal{RVV2}} & \sim & \varepsilon\bar\varepsilon^3\frac{(y_b y_t)}{9}^{0.5}\sin(\chi-2\beta'_2) \\
 (d_e)_{\textnormal{RVV3}} & \sim & \varepsilon\bar\varepsilon y_b y_t \sin(2(\chi-\delta_d))
\end{eqnarray}

Figure~\ref{fig:edms} shows the sensitivity of current~\cite{Regan:2002ta} and future~\cite{Lamoreaux:2001hb} $d_e$ experiments, for the three models. In order to be able to compare them consistently, all phases have been set to zero, except $\chi=\pi/4$. This maximizes $d_e$ in RVV1 and RVV3, and gives a large contribution to RVV2.

In the Figure, both RVV1 and RVV2 can be seen to survive the current constraints, but at the same time are large enough to be probed at future EDM experiments. In particular, for RVV1, the observation of a $d_e\sim10^{-29}$ would favour light SUSY masses. On the other hand, RVV2 predicts a value of $d_e$ of about one order of magnitude larger than RVV1 for any particular value of $m_0$ and $M_{1/2}$. This means that by reaching $d_e\sim10^{-29}$ one can probe a much larger part of the evaluated parameter space, with $m_0\lesssim1750$, $M_{1/2}\lesssim2000$.

For such phase configuration, RVV3 is greatly constrained by the current $d_e$ bounds. This means that, for SUSY masses with $m_0<1000$~GeV and $M_{1/2}\lesssim1200$~GeV, either $\chi$ must be supressed, or $\delta_d$ must have a value that cancels $\chi$. Such constraints seem unnatural and against the whole motivation for these kind of models. This leads one to conclude that both EDM and LFV data greatly disfavour RVV3 in most of the parameter space to be probed by the LHC, in contrast to RVV1 and RVV2.

\section{Conclusions}
It has been shown that it is possible to relate the SM and SUSY Flavour Problems using an $SU(3)$ family symmetry, and by doing so, both problems can be solved simultaneously. The breaking of $SU(3)$ can also lead to the breaking of exact CP, constraining all phases within the flavour sector, and solving the SUSY CP Problem.

Three possible exclusive variations of this model have been analyzed, and the observation of LFV processes, as well as EDMs, has proven to be essential to distinguish them. Of the three models, RVV1 and RVV2 survive the current LFV and EDM bounds even with light SUSY masses. If SUSY is observed at the LHC and one of these two models is at work, it would be very likely to observe $\mu\to e\gamma$ and $\tau\to\mu\gamma$ decays in the upcoming experiments, as well as an electron EDM, if the phases are not small. RVV3 turns out to be highly disfavoured by the low-energy observables in most of the evaluated parameter space.

Thus, the $SU(3)$ models turn out to have a testeable phenomenology in the short term. It is interesting to notice that, as the flavon phases in the lepton sector are the same as those of the quark sector, correlations between both sectors are expected to appear~\cite{WIP}. This would give more hints on the correct way of building the model, reducing in this way the ambiguity in the symmetries required for building the Yukawa and soft matrices.

\ack
I acknowledge support from the Spanish MCYT FPA2005-01678, and would like to thank O.~Vives and L.~Calibbi, who collaborated in the original work this talk is based on.

\section*{References}
\bibliography{discrete08.bib}{}

\end{document}